\documentclass[english,11pt]{article}
\pdfoutput=1
\usepackage{epsf,amsmath,amssymb,graphicx,dcolumn}
\usepackage{scalefnt,ulem}
\usepackage{subcaption}
\usepackage{cite,hyperref}
\usepackage{cleveref}
\usepackage{todonotes}
\usepackage{color}
\usepackage{rotating}
\usepackage{microtype}
\usepackage{listings} 
\usepackage{selinput} 
\usepackage{etoolbox}
\apptocmd{\thebibliography}{\setlength{\itemsep}{0.05cm}}{}{}
\definecolor{lightblue}{rgb}{.7,.8,1}
\Crefname{figure}{Fig.}{Figs.}
\title{\vspace*{-4em}
  \begin{flushright}
    {\sf\small
      DESY 16-146 --- KA-TP-24-2016 --- ZU-TH 27/16\\
    }
  \end{flushright}
\vspace*{2em}
Distributions for neutral Higgs production in the \nmssm{}\vspace{-0.2cm}}
\author{Stefan Liebler$^{a}$, Hendrik Mantler$^{b,c}$, Marius Wiesemann$^{d}$\\[1.2em]
{\it $^a$DESY, Notkestra{\ss}e 85, D-22607 Hamburg, Germany}\\[0.2em]
{\it ${}^b$ Institute for Theoretical Physics (ITP), Karlsruhe Institute of Technology,}\\[-0.2em]
{\it Engesserstra{\ss}e 7, D-76128 Karlsruhe, Germany}\\[0.2em]
{\it ${}^c$ Institute for Nuclear Physics (IKP), Karlsruhe Institute of Technology,}\\[-0.2em]
{\it Hermann-von-Helmholtz-Platz 1, D-76344 Eggenstein-Leopoldshafen, Germany}\\[0.2em]
{\it $^d$Physik-Institut, Universit\"at Z\"urich, CH-8057 Z\"urich, Switzerland}\\[0.2em]
{\small\tt stefan.liebler@desy.de}\\[-.3em]
{\small\tt hendrik.mantler@kit.edu}\\[-.3em]
{\small\tt mariusw@physik.uzh.ch}
}
\date{}

\newcommand{\abbrev}{\scalefont{.9}}
\newcommand{\mhiggs}{m_{\phi}}
\newcommand{\moresushi}{{\tt MoRe-SusHi}}
\newcommand{\amcsushi}{{\tt aMCSusHi}}
\newcommand{\powhegsushi}{{\tt POWHEG-SusHi}}

\newcommand{\powhegbox}{{\tt POWHEG BOX}}
\newcommand{\lhc}{{\abbrev LHC}}
\newcommand{\susy}{{\abbrev SUSY}}
\newcommand{\sushi}{{\tt SusHi}}
\newcommand{\bsm}{{\abbrev BSM}}        
\newcommand{\powheg}{{\abbrev POWHEG}}
\newcommand{\atl}{{\abbrev ATLAS}}
\newcommand{\cms}{{\abbrev CMS}}
\newcommand{\vev}{{\abbrev VEV}}
\newcommand{\pythia}{{\tt Pythia8}}
\newcommand{\pt}{\ensuremath{p_T}}
\newcommand{\nll}{{\abbrev NLL}}

\newcommand{\lo}{{\abbrev LO}}
\newcommand{\nlo}{{\abbrev NLO}}
\newcommand{\nnlo}{{\abbrev NNLO}}
\newcommand{\nnnlo}{{\abbrev N3LO}}
\newcommand{\sm}{{\abbrev SM}}
\newcommand{\qcd}{{\abbrev QCD}}
\newcommand{\thdm}{{\abbrev 2HDM}}
\newcommand{\nmssm}{{\abbrev NMSSM}}
\newcommand{\mssm}{{\abbrev MSSM}}
\newcommand{\ps}{{\abbrev PS}}
\newcommand{\plus}{{\abbrev +}}
\newcommand{\citere}[1]{Ref.\cite{#1}}
\newcommand{\citeres}[1]{Refs.\cite{#1}}
\newcommand{\eqn}[1]{Eq.\,(\ref{#1})}

\newcommand{\fig}[1]{Fig.\,\ref{#1}}

\newcommand{\sct}[1]{Section~\ref{#1}}

\newcommand{\muF}{\mu_{\rm F}}
\newcommand{\muR}{\mu_{\rm R}}
\newcommand{\pdf}{{\abbrev PDF}}
\newcommand{\mstw}{{\abbrev MSTW2008} $68${\abbrev \%CL}}
\newcommand{\madmc}{{\tt MadGraph5\_aMC@NLO}}
\newcommand{\mcatnlo}{{\abbrev MC@NLO}}
\newcommand{\cp}{{$\mathcal{CP}$}}

\def\RS{\mathcal{R}^S}
\def\RP{\mathcal{R}^P}
\def\RG{\mathcal{R}^G}

\newcounter{notecount}
\begin{document}
\maketitle

\begin{abstract}
\noindent
A novel computation of the fully-differential cross section for 
neutral Higgs-boson production through 
gluon fusion in the \cp{}-conserving \nmssm{} is presented. Based on the
calculation of \nlo{} corrections to the total cross section~\cite{Liebler:2015bka}, 
we implemented the \nmssm{}
amplitudes in three codes, applying different resummation techniques: analytic 
transverse-momentum resummation at \nlo{}\plus\nll{}, and 
two fully-differential \nlo{}\plus\ps{} Monte-Carlo approaches
using the \mcatnlo{} and \powheg{} matching procedures, respectively.
We study phenomenological predictions for distributions in the 
\nmssm{} with a special emphasis on the Higgs transverse-momentum 
spectrum. Reasonable agreement among the various approaches is 
found, once well-motivated choices for the unphysical matching scales and the determination 
of the related uncertainties are made.
\end{abstract}

\section{Introduction}
The discovery of a scalar resonance in searches for a Higgs boson by the
\atl{} \cite{Aad:2012tfa} and \cms{} \cite{Chatrchyan:2012xdj} collaborations is already considered 
as the legacy of Run~I of the Large Hadron Collider (\lhc{}). Although its measured properties 
are in full agreement with the Standard Model (\sm{}) predictions\footnote{See \citeres{Dittmaier:2011ti,Dittmaier:2012vm,Heinemeyer:2013tqa} for a theoretical overview.} so far, see e.g. \citeres{Aad:2015mxa,Khachatryan:2014jba},
the scalar particle may as well be embedded in an 
enlarged Higgs sector entailed in many beyond \sm{} (\bsm{}) theories.
Among the most relevant of such \sm{} extensions are supersymmetric models,
with the Minimal Supersymmetric
Standard Model (\mssm{}) being the simplest realization.
The latter can be further extended to the 
Next-to-Minimal Supersymmetric Standard Model (\nmssm{})
by adding an SU$(2)_L$ singlet. This has the advantage of a 
dynamical generation of the $\mu$-term~\cite{Ellwanger:2009dp,Maniatis:2009re}, 
which lifts the upper bound on the mass of the lightest Higgs boson at tree level. Hence, a 
\sm{}-like Higgs boson with a mass of $\sim125$\,GeV can be easily accommodated  in the \nmssm{}.

The \cp{}-conserving $\mathbb{Z}_3$-invariant \nmssm{} entails five neutral Higgs bosons
$\phi\in\lbrace H_1,H_2,H_3,A_1,A_2 \rbrace$, three of which are \cp{}-even and two 
\cp{}-odd. 
While in the \mssm{} the light Higgs boson is identified with the
observed \sm{}-like Higgs resonance, the \nmssm{} allows for phenomenologically 
interesting scenarios with
lighter (pseudo-)scalar mass eigenstates.
The direct search of further Higgs resonances is one of the central 
physics goals in \lhc{} Run~II. This requires precise theoretical
predictions in explicit models for both inclusive 
and differential observables.

The most important Higgs-production mechanism in the \sm{} is through gluon 
fusion. This remains true also in a large region of the parameter space 
of the {\abbrev (N)MSSM}, which, in fact, favors
lower values of $\tan\beta$ compared to the \mssm{}.
In these cases the Higgs-gluon coupling is predominantly mediated
by a top-quark loop.\footnote{The additional SU$(2)_L$ singlet of the \nmssm{} does not couple to quarks
and heavy gauge bosons directly. However, it couples directly to squarks.}
The total inclusive cross section has been computed in the limit of heavy top quarks at
next-to-next-to-leading order (\nnlo{}) 
\cite{Harlander:2002wh,Anastasiou:2002yz,Ravindran:2003um} and recently 
at next-to-\nnlo{} (\nnnlo{}) \cite{Anastasiou:2015ema,Anastasiou:2016cez}. 
Finite top-mass effects in the \sm{} have been estimated to be below $1\%$
at \nnlo{} \cite{Marzani:2008az,Harlander:2009bw,Harlander:2009mq,Harlander:2009my,
Pak:2009dg,Pak:2011hs}. One must bear in mind, however, that
in extended theories with two (or more) Higgs doublets the 
bottom-quark Yukawa coupling can be subject to a significantly enhancement,
so that, on the one hand, the bottom-quark as a mediator of the Higgs-gluon coupling 
becomes equally or even more important than the top-quark.\footnote{In such a case, the theoretical uncertainties for the gluon fusion
cross section can be substantial~\cite{Bagnaschi:2014zla}.} On the 
other hand, Higgs production in association with bottom quarks 
competes with the gluon-fusion process and can become the 
dominant production mode.\footnote{For Higgs production 
in association with bottom quarks in the four- and five-flavor schemes 
see \citeres{Campbell:2002zm,Harlander:2003ai,Dittmaier:2003ej,Dawson:2003kb,
Harlander:2010cz,Harlander:2011fx,Buehler:2012cu,Harlander:2014hya,Wiesemann:2014ioa} and references therein.}
The exact treatment of top and bottom-quark mass effects
\cite{Spira:1995rr,Harlander:2005rq} thus becomes vital for a precise 
prediction of Higgs cross sections in such theories.
Squark and gluino effects at \nlo{} \qcd{} are known in 
certain approximations. Their expansion in heavy
\susy{} masses in the \mssm{}, see \citeres{Degrassi:2010eu,Degrassi:2011vq,Degrassi:2012vt},
was adapted to the \nmssm{} in \citere{Liebler:2015bka}. All these effects 
are implemented in the numerical code \sushi{}\,\cite{Harlander:2012pb,Liebler:2015bka,Harlander:2016hcx}.

Kinematical distributions are an important tool in Higgs-boson 
measurements to distinguish the possible interplays between 
top- and bottom-quark, squark and gluino effects in the gluon-fusion 
scattering amplitude. This provides a precise discrimination between 
predictions in and beyond the \sm{}, and allows for the determination 
of exclusion limits on the parameter space of \bsm{} models. One of 
the most relevant differential observables in this respect is the 
Higgs transverse-momentum (\pt{}) spectrum. 
\nlo{}\,\cite{deFlorian:1999zd,Glosser:2002gm} and \nnlo{}\,\cite{Boughezal:2013uia,Chen:2014gva,Boughezal:2015dra,
Boughezal:2015aha,Caola:2015wna} \qcd{} predictions for this observable 
are known only in the limit of heavy top quarks. Top-quark mass 
effects in the \sm{} were found to be moderate ($\sim 2-3\%$) as long 
as momentum scales remain below roughly the top-quark mass or are integrated out 
\cite{Harlander:2012hf,Neumann:2014nha}.

Predictions valid at all transverse momenta require small-\pt{}
resummation of logarithmically enhanced terms 
to all orders in the strong coupling constant ($\alpha_s$). Such resummation can be derived 
from universal properties of \qcd{} radiation in the infra-red region
\cite{Dokshitzer:1978hw,Parisi:1979se,
Curci:1979bg,Collins:1981uk,Collins:1981va,Kodaira:1981nh,
Kodaira:1982az,Altarelli:1984pt,Collins:1984kg,Catani:2000vq}, and can be 
performed analytically\footnote{Another powerful technique to perform such resummation is soft-collinear effective
theory (SCET)~\cite{Bauer:2000ew,Bauer:2000yr,Bauer:2001ct,Bauer:2001yt,Beneke:2002ph}.} or 
by means of a numerical parton shower (\ps{}) approach. In the \mssm{}, 
the analytically resummed Higgs transverse-momentum spectrum in gluon fusion 
has been computed at next-to-leading logarithmic (\nll{}) accuracy with 
a consistent matching to the \nlo{} fixed-order cross section
\cite{Mantler:2012bj,Harlander:2014uea}, while fully-differential \nlo{} 
predictions matched to parton showers (\nlo\plus{}\ps{}) were first 
presented in the \powheg{} approach \cite{Nason:2004rx,Frixione:2007vw}
in \citere{Bagnaschi:2011tu}, and later in a \mcatnlo{}-type matching \cite{Frixione:2002ik}
in \citere{Mantler:2015vba}.

Common to all three approaches (analytic resummation, \mcatnlo{}, \powheg) 
is an effective matching scale (resummation scale, shower scale, $h_{\rm fact}$) 
that controls the separation between the soft/collinear and the hard region.
Although the dependence on the matching scales is of higher logarithmic order, 
suitable choices turn out to be absolutely crucial in particular 
when the bottom loop is involved. We refer the reader to 
\citeres{Harlander:2014uea,Bagnaschi:2015qta}, where two proposals for 
their algorithmic determination are made, and to \citere{Bagnaschi:2015bop} for 
the comprehensive comparison of these proposals.

In this letter, we present a novel computation of differential Higgs-boson production 
through gluon fusion in the \nmssm{} and report on new implementations 
of \nmssm{} effects in three different codes,
which all apply the {\tt SusHi} amplitudes
for the computation of the \nmssm{} matrix elements:\footnote{These 
codes feature \nlo{} accuracy (up to $\alpha_s^3$) on the total cross section,
implying formally only \lo{} accurate predictions at large transverse momenta.}
\begin{itemize}
\item \moresushi{} \cite{Mantler:2012bj,Harlander:2014uea,moresushiHP} performs analytic transverse-momentum resummation at \nlo{}\plus{}\nll{}.
\item \amcsushi{} \cite{Mantler:2015vba,amcsushiHP} 
employs the {\tt MadGraph5\_aMC@NLO} framework \cite{Alwall:2014hca}
to compute \nlo{}\plus\ps{} predictions with the \mcatnlo{} method.\footnote{\amcsushi{} has also been applied for the 
lowest multiplicities in the recent computation of Higgs production with
multi-jet merging up to two jets at \nlo{}\plus{}\ps{} \cite{Frederix:2016cnl}.}
\item \powhegsushi{} \cite{Mantler:XX} uses the corresponding \powheg{} implementation for the 
\nlo{}\plus\ps{} matching in the \powhegbox{} framework \cite{Alioli:2010xd}.
\end{itemize}

\section{The Higgs sector of the \nmssm{}}
\label{sec:nmssm}

We start with a discussion of the Higgs sector of the \cp-conserving $\mathbb{Z}_3$-invariant \nmssm{},
which closely follows \citeres{Ender:2011qh,Liebler:2015bka}. If we denote the superpotential of
the \mssm{} (without $\mu$-term) by $W_{\rm\mssm}$, the one of the \nmssm{} is given by
\begin{align}
W_{\rm\nmssm} = W_{\rm\mssm}
-\epsilon_{ab} \lambda\hat{S}\hat{H}_d^a\hat{H}_u^b + \frac{1}{3}\kappa \hat{S}^3\,,
\end{align}
where $\hat{H}_d$ and $\hat{H}_u$ are the two SU$(2)_L$ doublet superfields known from
the \mssm{} and $\hat{S}$ is the additional SU$(2)_L$ singlet superfield.
The tensor $\epsilon_{ab}$ contracts the components of the SU$(2)_L$ doublets.
The singlet $\hat{S}$ is a neutral superfield and thus the \nmssm{} contains one
additional \cp{}-even and \cp{}-odd (two and three in total) neutral Higgs boson  compared to the \mssm{}.
The soft-breaking terms can be written in the form
\begin{align}
\label{eq:softlag}
 \mathcal{L}_{\text{soft}} = \mathcal{L}_{\text{soft},\mssm}
 + (\epsilon_{ab}\lambda A_\lambda S H_d^a H_u^b
 - \frac{1}{3}\kappa A_\kappa S^3 + \text{h.c.}) - m_s^2|S|^2\,.
\end{align}
Similar to $m_{H_d}^2$ and $m_{H_u}^2$ in the \mssm{}, the mass term $m_s^2$
can be obtained from the minimization conditions of the
tadpole equations.
We consider $A_\lambda$ and $A_\kappa$ as input parameters. Alternatively,
$A_\lambda$ can be replaced by the mass~$m_{H^\pm}$ of the charged Higgs boson
as input parameter. 
We decompose the neutral components of the Higgs fields as follows
\begin{align}
 H_u^0=\frac{1}{\sqrt{2}} (v_u + H_u^R + iH_u^I)\,,\quad
 H_d^0=\frac{1}{\sqrt{2}} (v_d + H_d^R + iH_d^I)\,,\quad
  S= \frac{1}{\sqrt{2}}   (v_s + S^R + iS^I)\,,
\end{align}
where $v_d,v_u$ and $v_s$ are the vacuum expectation values (\vev{}s) 
and the fields with indices $R$ and $I$ denote the \cp{}-even and -odd fluctuations
around the \vev{}s. The \vev{} of the singlet is $v_s$ and generates an effective 
$\mu$-term $\mu_{\rm eff} = \lambda v_s/\sqrt{2}$, which is considered also as input parameter.
Consequently $v_s$ is determined by $\lambda$ and $\mu_{\rm eff}$.
We refrain from a detailed discussion of the mass matrices, 
but rather refer to \citere{Ender:2011qh}. The mixing matrices,
transforming the gauge into mass eigenstates, are crucial though. 
For this purpose we define the
basis of \cp{}-even gauge eigenstates $H^R=(H_d^R,H_u^R,S^R)$ and
the one of the \cp{}-odd eigenstates $H^I=(H_d^I,H_u^I,S^I$).
In this notation
the mass eigenstates $H_i$ with $i\in\lbrace 1,2,3\rbrace$ in the \cp{}-even case
are easily obtained through a single rotation with a $(3\times 3)$-matrix
\begin{align}
H_i = \sum_{j=1}^3 \RS_{ij} H^R_j\,.
\end{align}

In the \cp{}-odd sector we perform a prerotation $\RG$
such that the intermediate basis is given by $H'^I=(A,S^I,G)$ with
the Goldstone boson as last entry.\footnote{We follow the convention of \sushi{} release {\tt 1.6.0}, 
being different from the notation of \citere{Liebler:2015bka} and \sushi{} {\tt 1.5.0}.}
The final mass eigenstates $A_i$ with $i\in\lbrace 1,2\rbrace$ are obtained through
\begin{align}
A_{i} = \sum_{j=1}^3\RP_{ij}H'^I_j
=\sum_{j,k=1}^3\RP_{ij}\RG_{jk}H^I_k \quad \text{with} \quad 
\RG=\begin{pmatrix}s_\beta&c_\beta&0\\0&0&1\\c_\beta&-s_\beta&0\end{pmatrix}\,.
\end{align}
The $(3\times 3)$-matrix $\RP$ consists of a $(2\times 2)$-mixing block,
whereas $\RP_{i3}=\RP_{3i}=0$ for $i\neq 3$ and $\RP_{33}=1$.
The mass spectrum and Higgs mixing matrices can thus be easily taken over
from various spectrum generators, see \citere{Staub:2015aea} and references therein.

For the subsequent study
we pick a benchmark scenario (named BP2\_P2) provided by the \lhc{} Higgs Cross Section Working
Group \cite{benchmarkLHCHXSWG}, which
is based on \citeres{Bomark:2014gya,Bomark:2015fga} and defined by
\begin{align}
\tan\beta=2.266,\; \lambda=0.644,\; \kappa=0.351,\; \mu_{\rm eff}=178~\text{GeV},\; A_{\kappa}=100~\text{GeV},\; A_{\lambda} = -312~\text{GeV}\,.
\end{align}
We use {\tt NMSSMTools 4.7.0} \cite{Ellwanger:2004xm,Ellwanger:2005dv,Belanger:2005kh,Ellwanger:2006rn} to obtain the Higgs spectrum with masses
\begin{align}\begin{split}
 &m_{H_1}=125.9~\text{GeV}, \quad m_{H_2}=201.0~\text{GeV}, \quad m_{H_3}=448.1~\text{GeV}\\
 &m_{A_1}=65.2~\text{GeV}, \quad m_{A_2}=440.0~\text{GeV}\,.
\end{split}\end{align}
The input parameters of the relevant colored SUSY particles are given by
\begin{align}\begin{split}
 &m_{\tilde{g}}=2222~\text{GeV}, \quad A_t=-1630~\text{GeV}, \quad A_b=-3376~\text{GeV},\\
 &m_{Q}=1710~\text{GeV}, \quad m_{U}=1046~\text{GeV}, \quad m_{D}=2064~\text{GeV}\,,
\end{split}\end{align}
where $m_Q, m_U, m_D, A_t$ and $A_b$ are the soft-breaking parameters
of the third-generation squark sector in $\mathcal{L}_{\text{soft},\mssm}$ of \eqn{eq:softlag}.
The light scalar in that scenario is \sm{}-like and thus consistent with 
data of \lhc{} Run I. Furthermore, the low-mass pseudo-scalar is compatible 
with \lhc{} and LEP searches. We note that the scenario features heavy
squarks and gluinos, with on-shell masses above $1$\,TeV,
lowering their impact on the gluon-fusion cross section.

We subsequently discuss differential results 
for the two singlet-like Higgs bosons $A_1$ and $H_2$ as well as one standard heavy Higgs boson $H_3$.
By contrast, we refrain from discussing the \sm{}-like Higgs boson $H_1$ as well as the \mssm{}-like pseudoscalar~$A_2$
as they do not provide new features.

We also do not consider a benchmark scenario with a singlet-like (pseudo-)scalar, for which
quark-contributions tend to cancel each other and squark as well as electro-weak
contributions are accordingly large, see \citere{Liebler:2015bka} for an example,
since the cross section in such a case is rather small 
and the experimental determination of the transverse-momentum spectrum remains very 
challenging even in \lhc{} Run~II.

\section{Differential cross sections in the \nmssm{}}
\label{sec:results}

\begin{figure}[t]
  \begin{center}
    \begin{tabular}{ccccccc}
      \mbox{\includegraphics[height=.09\textheight]{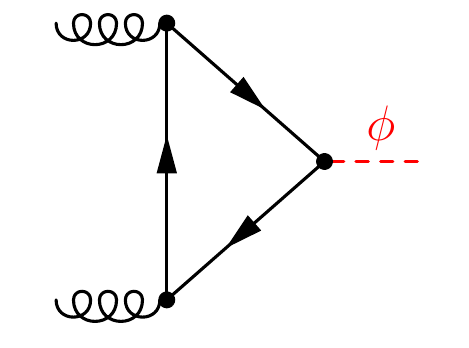}} & & \mbox{\includegraphics[height=.09\textheight]{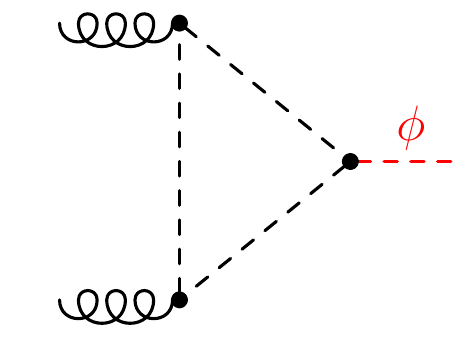}} & &
      \mbox{\includegraphics[height=.09\textheight]{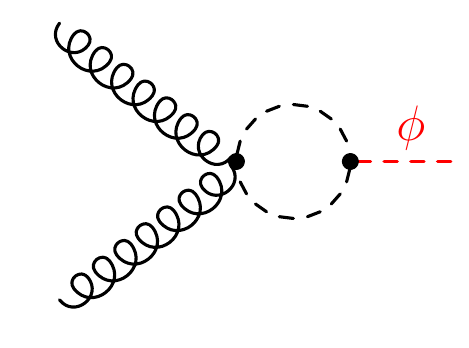}} & & 
      \mbox{\includegraphics[height=.09\textheight]{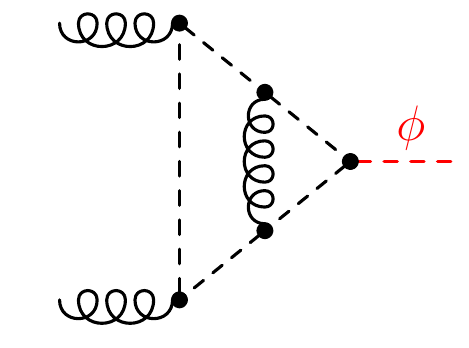}} \\
      (a) & & (b) & & (c) && (d)\\[0.3cm]
\mbox{\includegraphics[height=.09\textheight]{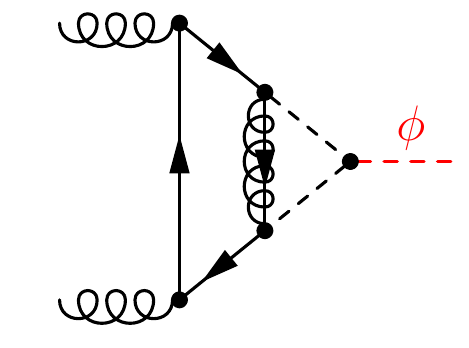}} & &
      \mbox{\includegraphics[height=.09\textheight]{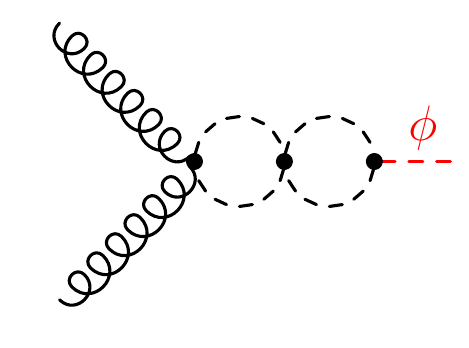}} & & \mbox{\includegraphics[height=.065\textheight]{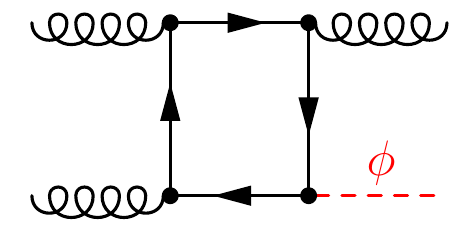}} & &
      \mbox{\includegraphics[height=.085\textheight]{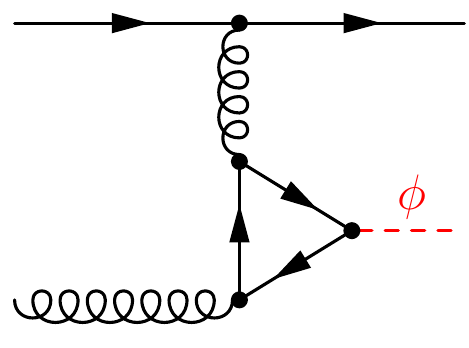}} \\
      (e) & & (f) & & (g) && (h)
    \end{tabular}
      \caption[]{\label{fig:diag} A sample of Feynman diagrams for
        $gg\rightarrow \phi$ contributing to the \nlo{} cross section;
        (a-c) \lo{}, (d-f) virtual and (g-h) real corrections. The
        graphical notation for the lines is: solid straight
        $\widehat{=}$ quark; spiraled $\widehat{=}$ gluon; dashed
        $\widehat{=}$ scalar (squark or Higgs); spiraled with line
        $\widehat{=}$ gluino.  }
  \end{center}
\end{figure}

The generic leading-order diagrams for Higgs production in 
the \nmssm{}, i.e., $gg\rightarrow\phi$ with $\phi\in\{H_1,H_2,H_3,A_1,A_2\}$, 
are induced by a quark or 
squark loop, as shown in \fig{fig:diag}\,(a-c). At \nlo{} 
each of these diagrams receives virtual corrections by a gluon 
or gluino exchange, 
e.g., \fig{fig:diag}\,(d-e), or by an additional squark loop,
e.g., \fig{fig:diag}\,(f). Additionally, real emission 
contributions must be included, which involve also diagrams with  
further initial-state partons, see \fig{fig:diag}\,(g-h) for example. 
The corresponding \nmssm{} matrix elements are implemented in and taken from 
the code \sushi{}\,\cite{Liebler:2015bka,Harlander:2012pb,Harlander:2016hcx}. 

They are combined and matched to all-order results in three 
different resummation frameworks. The analytic resummation 
of soft and collinear logarithms as formulated in \citere{Collins:1984kg} 
with the matching procedure developed in \citere{Bozzi:2005wk}
allows for the prediction of a single differential observable, 
the inclusive transverse-momentum spectrum of the Higgs boson, but with the highest 
possible order in the logarithmic series, i.e., \nll{} 
accuracy in the {\abbrev (N)MSSM}. Monte-Carlo approaches perform resummation numerically 
by means of a parton shower. The matching to the \nlo{} cross section is 
done by the well-known \mcatnlo{}\,\cite{Frixione:2002ik} and 
\powheg{}\,\cite{Nason:2004rx,Frixione:2007vw} methods. They allow 
for the computation of arbitrary infra-red safe observables 
at \nlo\plus{}\ps{}, but consistently resum only the leading 
logarithms and partially the ones beyond. For the Monte-Carlo simulations we 
use the framework of \madmc{} and \powhegbox{}, respectively, which 
support all standard parton showers 
\cite{Corcella:2000bw,Corcella:2002jc,Sjostrand:2006za,%
Sjostrand:2007gs,Bahr:2008pv,Bellm:2013lba}.

All phenomenological results in this letter are obtained with the codes 
\moresushi{}, \amcsushi{} and \powhegsushi{}.\footnote{Note that 
due to the application of the same matrix elements,
the three codes work at the same perturbative accuracy, i.e., \nlo{} 
\qcd{} (up to $\alpha_s^3$) regarding the total cross section.} 
In particular, we employ the modified \powheg{} (m\powheg{}) matching to 
the shower as suggested in \citere{Bagnaschi:2015bop} for all the \powhegsushi{} results, 
by restricting the shower starting scale to a fixed value instead of using 
the transverse momentum of the first emission, which may become arbitrarily 
large. This has the advantage of limiting the impact of the parton shower at large 
transverse momenta, where it is outside its validity region and the fixed-order 
result gives a viable prediction, and warrants a smooth merging of the matched result
into the fixed-order cross section at large transverse momenta.

We work at the $13$\,TeV \lhc{} and use the \mstw{} \nlo{} \pdf{} sets \cite{Martin:2009iq}, 
with the associated value of the strong coupling constant.\footnote{Other
\pdf{} sets supported by the {\tt LHAPDF}~\cite{Buckley:2014ana} interface can be employed in all three codes.}
Top and bottom-quark masses are renormalized on-shell both in the matrix 
elements and for the Yukawa couplings, and set to the pole masses 
$m_t=172.5$\,GeV and $m_b = 4.75$\,GeV, respectively.
We use dynamical central renormalization and factorization scales of 
$\muR=\muF=H_T/2\equiv1/2\,\sum_{i}(m_i^2+p_T^2(i))^{1/2}$ ($i$ runs over 
all final-state particles, with $m_i$ and $p_T(i)$ being their respective mass and 
transverse momentum) for the 
Monte-Carlo predictions and $\muR=\muF=m_T/2\equiv1/2\,(\mhiggs^2+\pt^2(\phi))^{1/2}$
in analytic resummation and the fixed order distribution (f\nlo{}). 
The matching scales in the various approaches (resummation scale, shower scale, 
$h_{\rm fact}$) are chosen as suggested in \citere{Harlander:2014uea,Bagnaschi:2015bop}, see also \citere{Mantler:2015vba} regarding the implications for 
the dynamical shower scale choice in \amcsushi{}.
As far as the Monte-Carlo approaches are concerned, we use the {\tt Pythia8} \cite{Sjostrand:2007gs} parton shower.

\begin{figure}[t]
\begin{center}
	\begin{subfigure}[b]{0.48\textwidth}
         	\mbox{\includegraphics[width=\textwidth,page=1,trim={3.5cm 0.2cm 2.4cm 0.7cm}]{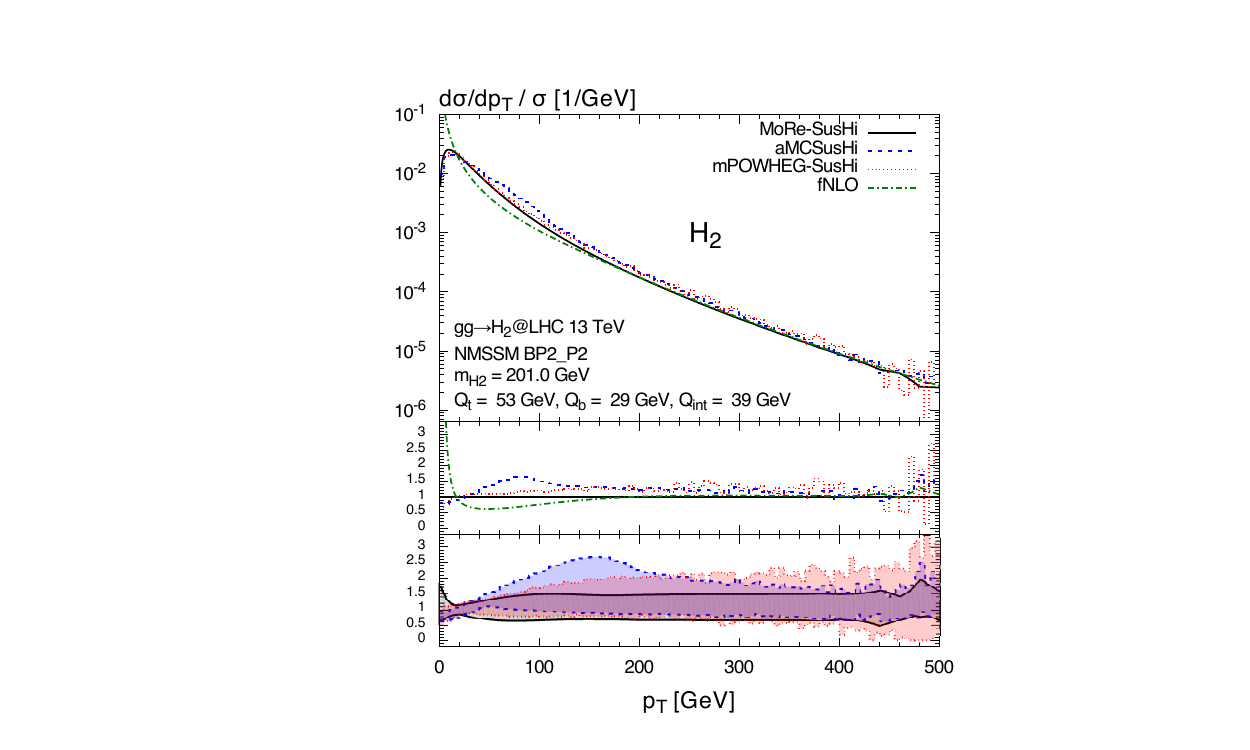}}
	        \caption{}
            \label{fig:veto_NLO_mH_abs}
        \end{subfigure}%
	\begin{subfigure}[b]{0.48\textwidth}
         	\mbox{\includegraphics[width=\textwidth,page=2,trim={3.5cm 0.2cm 2.4cm 0.7cm}]{plots/HT2.pdf}}
		\caption{}
        \label{fig:veto_NLO_mH_rel}
        \end{subfigure}%
\begin{center}\vspace{-0.4cm}
        \caption{\label{fig:H2H3} Transverse-momentum distributions of the two heavy scalars of the \nmssm{} scenario 
        introduced in \sct{sec:nmssm}; (a) $H_2$; and (b) $H_3$. See text for the description of the different curves.
        } \end{center}
\end{center}
\end{figure}

In \fig{fig:H2H3} we compare the predictions for the inclusive transverse-momentum distributions of the 
heavy scalar Higgs bosons in the \nmssm{} benchmark scenario BP2\_P2 (see \sct{sec:nmssm}) 
among the various approximations: \nlo{}\plus{}\nll{} analytic resummation (black, solid), \nlo{}\plus{}\ps{} 
in the \mcatnlo{} (blue, dashed), and the m\powheg{} (red, dotted) and, for reference, f\nlo{} (green, dash-dotted)
which formally corresponds to \lo{} as far as 
the \pt{} distribution is concerned. The f\nlo{} curve is computed with the same central scales as the analytically resummed result; 
we checked, however, that there are only minor differences to the scale adopted in the two Monte Carlos and only at very large \pt{}.
The bands correspond to a 7-point $(\muR{}$, $\muF{})$ variation by a factor of two around the central scale.
The variation of the matching scale in each approach is done also by a factor of two and added in quadrature.
In the case of analytic resummation, we apply, however, a suppression factor to the resummation-scale 
dependence, as introduced in \citere{Harlander:2014uea}, at large transverse momenta, since any dependence 
of the cross section on its value is necessarily artificial in the large-\pt{} region.

The left and right panels of \fig{fig:H2H3} correspond to the results for $H_2$ with $m_{H_2}=201.0$\,GeV and $H_3$ 
with $m_{H_3}=448.1$\,GeV, respectively. The features of the two plots are fairly similar, which allows us to discuss 
them simultaneously. The general observations are the following:
\begin{itemize}
\item Overall, the agreement among the different codes is reasonably well within the respective uncertainties, which 
are generally large though, in particular at high transverse momenta.
\item The analytically resummed curve of \moresushi{} is softer than the Monte-Carlo results. The two Monte-Carlo 
results are hardly distinguishable at very small transverse momenta ($0$\,GeV$\le\pt\le40$\,GeV), which is driven by 
the underlying \pythia{} shower. Also, the difference to \moresushi{} in that region is not very big either, reaching up to 
$\mathcal{O}(30$-$40\%)$ only in the first bin.
\item In the intermediate-\pt{} region ($50$\,GeV$\lesssim\pt\lesssim250$\,GeV) the \mcatnlo{} prediction of \amcsushi{} develops a somewhat larger uncertainty band than the other two approaches. As shown in \citere{Bagnaschi:2015bop} the main 
reason for this is the dynamical shower scale adopted from {\tt MadGraph5\_aMC@NLO}. Within that region (around 
$\pt{}\sim 100$\,GeV) also the central \amcsushi{} prediction deviates from the other results by $\mathcal{O}(50\%)$, 
which in turn are in rather well agreement, in particular regarding their shapes.
\item At large transverse momenta ($\pt\ge250$\,GeV), both Monte-Carlo predictions feature uncertainties that are 
larger than the ones of \moresushi{}, the \powheg{} band being particularly sizable in that region.
One should bear in mind, however, that for \moresushi{} we turned off 
uncertainties related to the resummation scale in that \pt{} region. 
We further note that by construction both \mcatnlo{} and m\powheg{}
results will eventually merge into the fixed-order result at sufficiently large transverse momenta also with respect 
to the scale uncertainties. Indeed, we already see the smooth matching to the fixed-order curve at large \pt{} for the 
central curves. Though small differences will remain due to the slightly different choices adopted for the 
central scales of f\nlo{} and the two Monte Carlos.
\item We finally remark that the general differences we observe among the codes are very reminiscent to what 
has been found in \citere{Bagnaschi:2015bop} in case of generic \thdm{} Higgs bosons. The general conclusions drawn in that paper are 
thus also applicable to the \nmssm{}, since the paper captures different hierarchies between top- and bottom-Yukawa couplings. 
\end{itemize}

\begin{figure}[t]
	\begin{center}
	   \mbox{\includegraphics[width=0.48\textwidth,page=3,trim={3.6cm 0.15cm 2.3cm 0.7cm}]{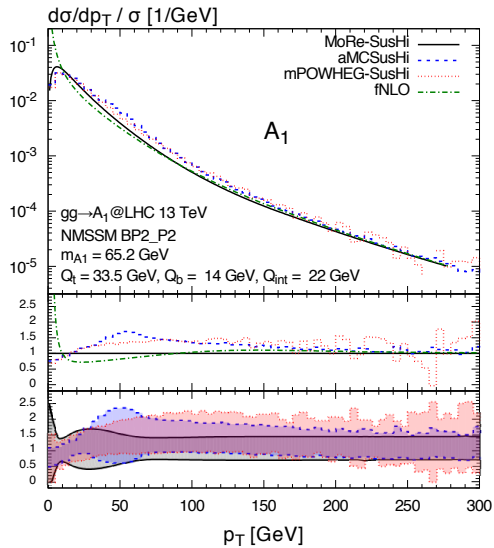}}
        \caption{\label{fig:A1} Same as \fig{fig:H2H3}, but for the light pseudo-scalar Higgs boson $A_1$.
                } \end{center}
\end{figure}

We now turn to the transverse-momentum distribution of the light pseudo-scalar 
Higgs boson ($A_1$) with $m_{A_1}=65.2$\,GeV, shown in \fig{fig:A1}. The general 
features are not very different as compared to the heavy scalars in \fig{fig:H2H3}. Most 
remarkable, however, is the excellent agreement between the central predictions of the two 
Monte Codes in this case, except for the small bump of the \amcsushi{} curve around 
$\pt\sim40-70$\,GeV. Similar to the scalar Higgs case also for the light pseudo-scalar particle 
the uncertainty in the \mcatnlo{} matching blows up, although to a lesser extend, and the \powheg{} 
curve develops a wider uncertainty band in the high-\pt{} region. 
The analytically resummed result of \moresushi{} again shows a generally softer 
spectrum, most apparent at small transverse momenta, while the uncertainties in that region 
are quite similar to the ones of the Monte Carlos, with a somewhat larger band though as $\pt\rightarrow 0$. 
The smaller band of \moresushi{} at high \pt{} is again caused by the suppression of the 
resummation-scale variations, although one already observes also for the \amcsushi{} prediction the smooth 
merging into the fixed-order result. Overall, the agreement of the various predictions is well within 
their respective uncertainties. We finally note that care must be taken, when considering very low Higgs boson 
masses ($m_\phi\lesssim 20$\,GeV) in computations such as the one at hand, since the Higgs
mass approaches quark thresholds, e.g. $2m_b$, which requires the resummation of
gluon effects.

\section{Conclusions}
\label{sec:conclusions}

We have presented predictions for Higgs-boson production through gluon fusion
in the \nmssm{}. For the first time tools are made available to simulate a Higgs 
signal at \nlo{} \qcd{} accuracy matched to parton showers and to compute the Higgs 
transverse-momentum distribution analytically at \nlo{}\plus{}\nll{}. We have 
considered the transverse-momentum distribution of various Higgs bosons in a 
\nmssm{} benchmark scenario, including the phenomenologically interesting case 
of a light pseudo-scalar of mass well below the observed scalar resonance.
The comparison of the predictions obtained with the three codes turned out to 
show rather similar features as the ones already observed in the \mssm{}. Overall, 
the agreement among these predictions is well within the theoretical 
uncertainty bands. Each of these codes, therefore, provides a proper modelling of 
the Higgs transverse-momentum spectrum, as long as the relevant (perturbative 
and resummation-related) scale uncertainties are kept into account.

The computations presented in this letter 
enable a reliable simulation of a Higgs signal within the 
\nmssm{} and will serve particularly useful in this respect to light and  heavy
Higgs-boson 
searches by the experiments at the \lhc{}. The combination of the gluon-induced 
cross-section predictions with the ones obtained for 
Higgs-boson production in association with 
bottom quarks, which becomes relevant in scenarios with an enhanced bottom-quark 
Yukawa coupling, was not discussed in this letter. However, Higgs-boson production
in association with bottom quarks and its distributions can be reweighted
from the corresponding \sm{} results and incoherently added to the gluon-fusion cross section.
Finally, off-shell effects become relevant 
for Higgs-boson masses around (and above) the TeV scale; their inclusion in the codes 
at hand is feasible, but beyond the scope of this letter.

\paragraph{Acknowledgements.}
We would like to thank the \lhc{} Higgs Cross Section Working Group for providing the motivation to perform this computation.
This research was supported in part by the Swiss National Science Foundation (SNF) under contract 200021-156585 and
by Deutsche Forschungsgemeinschaft through the SFB~676 ``Particles, Strings, and the Early Universe''.

\linespread{0}\selectfont
\bibliographystyle{h-physrev5}
\bibliography{ggh_bib}

\end{document}